\documentclass{article}
\usepackage{amsmath,graphicx,mlspconf}

\usepackage{cite}
\usepackage{comment}
\usepackage{amssymb}
\usepackage{color}
\usepackage{mathtools}
\usepackage{enumitem}
\usepackage{multirow}
\usepackage[table,xcdraw]{xcolor}
\usepackage{algorithm} 
\usepackage{algorithmic}
\usepackage[algo2e]{algorithm2e}

\newcommand{\etal}{\textit{et al}. }
\newcommand{\ie}{\textit{i}.\textit{e}., }

\setlist{nolistsep}

\DeclareMathOperator{\diag}{\operatorname{diag}}
\DeclareMathOperator{\vectorized}{\operatorname{vec}}

\DeclareMathOperator{\tr}{\operatorname{tr}}

\newtheorem{theorem}{Theorem}

\newtheorem{definition}{Definition}

\title{On the Minimization of Sobolev Norms of Time-Varying Graph Signals: Estimation of New Coronavirus Disease 2019 Cases}
\name{Jhony H. Giraldo and Thierry Bouwmans}
\address{Laboratoire MIA, La Rochelle Université, France\\
\{jgiral01, tbouwman\}@univ-lr.fr}

\begin{document}

\maketitle

\begin{abstract}
The mathematical modeling of infectious diseases is a fundamental research field for the planning of strategies to contain outbreaks. The models associated with this field of study usually have exponential prior assumptions in the number of new cases, while the exploration of spatial data has been little analyzed in these models. In this paper, we model the number of new cases of the Coronavirus Disease 2019 (COVID-19) as a problem of reconstruction of time-varying graph signals. To this end, we proposed a new method based on the minimization of the Sobolev norm in graph signal processing. Our method outperforms state-of-the-art algorithms in two COVID-19 databases provided by Johns Hopkins University. In the same way, we prove the benefits of the convergence rate of the Sobolev reconstruction method by relying on the condition number of the Hessian associated with the underlying optimization problem of our method.
\end{abstract}
\section{Introduction}
\label{sec:introduction}

The Coronavirus Disease 2019 (COVID-19) brought an unprecedented sanitarian crisis around the world in 2020 \cite{xu2020pathological} with a pandemic. Several developed countries such as the United States of America (USA), Italy, Spain, France, the United Kingdom, among others, have had problems trying to contain the outbreak. The capacity of some countries to detect new cases has been overwhelmed by the exponential number of cases, leading to a poor and unreliable estimation of the number of new cases. As a matter of fact, Colombia slowed down the testing of suspected new cases of COVID-19 between March 25th and 27th in 2020 because of a failure in an essential machine for the COVID-19 diagnosis.

The mathematical modeling of infectious diseases is an old and well-established research field \cite{daley2001epidemic,hethcote2000mathematics}, these infectious models are either stochastic or deterministic and use basic assumptions. Specifically in COVID-19, Wang \etal \cite{wang2020phase} used a Susceptible, Exposed, Infectious, and Removed (SEIR) model to estimate the epidemic trend in Wuhan, China. However, this model does not take into account the underlying spatial information of the problem, \ie the model does not associate the number of new cases in nearby localities. In the same way, not all countries are able to effectively sample the number of infected people in their population.

In this work, we model the number of new COVID-19 cases as a reconstruction of time-varying graph signals \cite{shuman2013emerging,sandryhaila2013discrete,ortega2018graph}. Our algorithm associates a graph to the geographical localization of cities or countries, and the number of new COVID-19 cases is represented by a graph signal that evolves in time \cite{loukas2016frequency,perraudin2017towards,qiu2017time}. Intuitively, the prior assumption of this work is that the number of new COVID-19 cases is smooth both in the graph and in time, \ie the number of new COVID-19 cases should be similar in nearby localities, and the change in time is progressive. Unlike previous mathematical models of infectious diseases, this paper takes into account the underlying spatial information of the problem. In the same way, we propose a new time-varying graph signals reconstruction algorithm based on the minimization of the Sobolev norm of Graph Signal Processing (GSP) \cite{pesenson2009variational}. Our method outperforms state-of-the-art algorithms in time-varying graph signals, while showing convergence rate benefits based on the Hessian of the underlying problem. The main contributions of this paper are summarized as follows:

\begin{itemize}
    \item To the best of our knowledge, concepts of GSP are introduced for the first time in the domain of infectious diseases modeling.
    \item A new algorithm based on the minimization of the Sobolev norm is introduced in the problem of time-varying graph signals reconstruction.
    \item We show the convergence rate benefits on the minimization of the Sobolev norm for the reconstruction of time-varying graph signals.
\end{itemize}


The rest of the paper is organized as follows. Section \ref{sec:algorithm} explains the basic concepts and the proposed method. Section \ref{sec:experimental_framework} introduces the experimental framework. Finally, Sections \ref{sec:results_and_discussion} and \ref{sec:conclusions} present the results and conclusions, respectively.

\section{Time-Varying Graph Signals Reconstruction}
\label{sec:algorithm}

\begin{figure}
    \centering
    \includegraphics[width=0.48\textwidth]{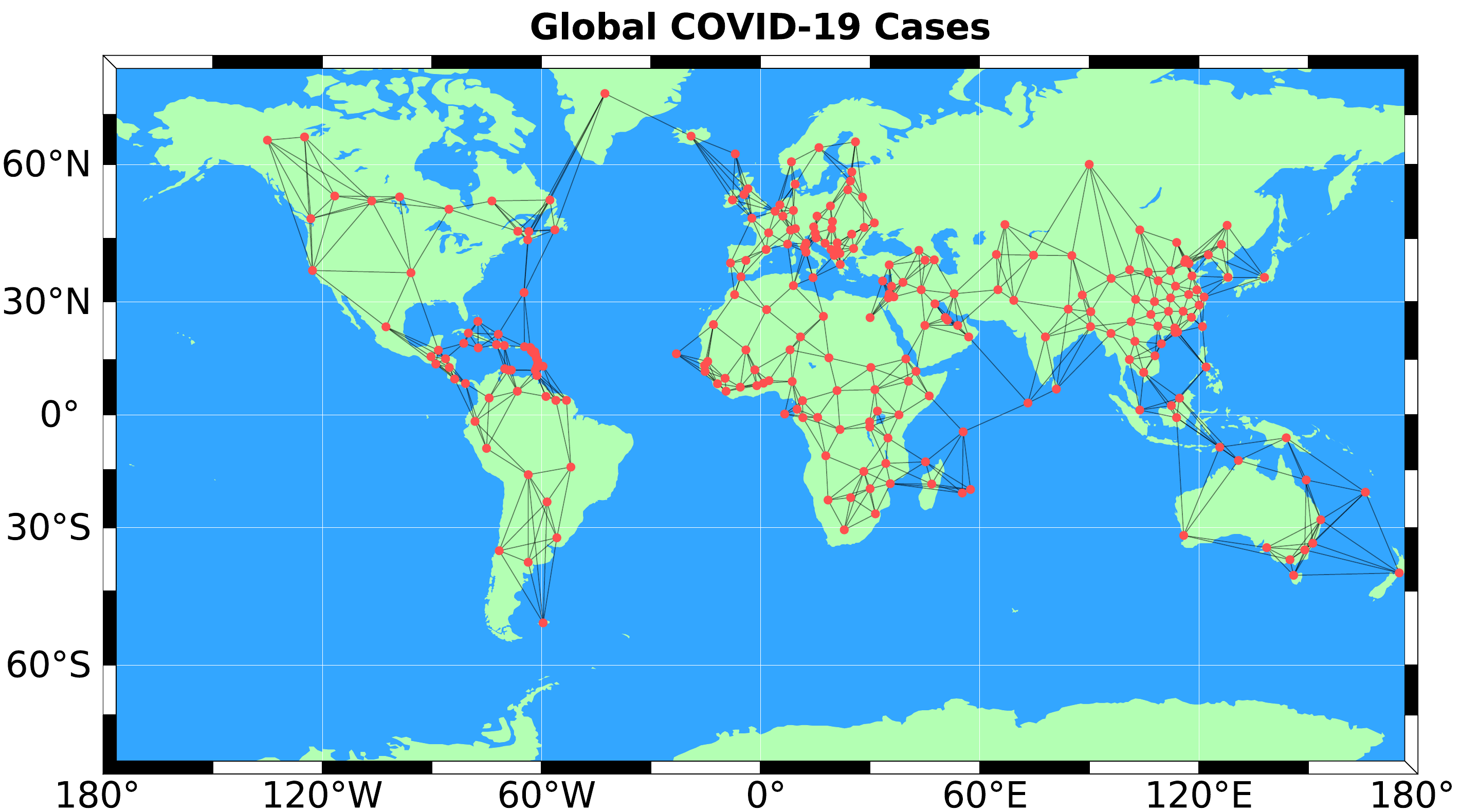}
    \caption{Graph with the regions in the world with confirmed cases of COVID-19 by April 6, 2020. This graph was constructed with k-nearest neighbors with $\text{k}=5$.}
    \label{fig:global_covid_cases}
\end{figure}

This section presents the mathematical notation, basic concepts of this paper, as well as the proposed Sobolev norm reconstruction algorithm. Figure \ref{fig:global_covid_cases} shows a graph with the regions in the world with confirmed cases of COVID-19 by April 6, 2020.

\subsection{Notation}
\label{sec:notation}

In this paper, uppercase boldface letters such as $\mathbf{X}$ represent matrices, and lowercase boldface letters such as $\mathbf{x}$ denote vectors. Calligraphic letters such as $\mathcal{E}$ represent sets. The Hadamard and Kronecker products between matrices are denoted by $\circ$ and $\otimes$, respectively. $(\cdot)^{\mathsf{T}}$ denotes transposition. The vectorization of a matrix $\mathbf{A}$ is denoted as $\vectorized{(\mathbf{A})}$, while $\diag(\mathbf{x})$ is a diagonal matrix with entries $x_1,x_2,\dots,x_n$. The trace and Frobenius norm of a matrix are represented by $\tr(\cdot)$ and $\Vert \cdot \Vert_F$, respectively.

\subsection{Background}

Let $G=(\mathcal{V,E})$ be a graph with $\mathcal{V}=\{1,\dots,N\}$ the set of $N$ nodes. $\mathcal{E}=\{(i,j)\}$ represents the set of edges, where $(i,j)$ is an edge between the nodes $i$ and $j$. $\mathbf{W} \in \mathbb{R}^{N\times N}$ is the weighted adjacency matrix of $G$, with $\mathbf{W}(i,j)>0 \Leftrightarrow \{i,j\} \in \mathcal{E}$. This paper is focused in undirected graphs, then $\mathbf{W}$ is symmetric, \ie the edges $(i,j)$ and $(j,i)$ have the same weight. $\mathbf{D} \in \mathbb{R}^{N\times N}$ is a diagonal matrix such that $\mathbf{D}(i,i)=\sum_{j=1}^N \mathbf{W}(i,j)~\forall~i = 1,\dots,N$. Furthermore, $\mathbf{L} = \mathbf{D-W}$ is the positive semi-definite combinatorial Laplacian operator of $G$, with eigenvalues $0=\lambda_1 \leq \lambda_2 \leq \dots \leq \lambda_N$ and corresponding eigenvectors $\{ \mathbf{u}_1,\mathbf{u}_2,\dots,\mathbf{u}_N\}$. And finally, a graph signal is a function in $\mathcal{V}$ such that $x:\mathcal{V} \to \mathbb{R}$, and it can be represented as a vector $\mathbf{x} \in \mathbb{R}^N$ where $\mathbf{x}(i)$ is the value of the function in the node $i \in \mathcal{V}$.

The sampling and reconstruction of graph signals play a central role in GSP \cite{chen2015discrete,anis2016efficient,romero2016kernel,parada2019blue}. Several algorithms for sampling and recovery assume that the graph signal is smooth in the graph. One well-known measure of smoothness in $G$ is the graph Laplacian quadratic form defined as $S_2(\mathbf{x})=\mathbf{x}^{\mathsf{T}}\mathbf{Lx}$. For example in reconstruction, Puy \etal \cite{puy2018random} used this Laplacian quadratic form as a regularization term in the formulation of their optimization problem. However, the graph Laplacian quadratic form $S_2(\mathbf{x})$ is limited to static graph signals in $G$. Qiu \etal \cite{qiu2017time} extended the definition of $S_2$ to time-varying graph signals. Let $\mathbf{X} = [\mathbf{x}_1, \mathbf{x}_2, \dots, \mathbf{x}_M]$ be a time-varying graph signal, where $\mathbf{x}_t$ is a graph signal in $G$ at time $t$. The smoothness of time-varying graph signals $\mathbf{X}$ is such that \cite{qiu2017time}:
\begin{equation}
    S_2(\mathbf{X}) = \tr(\mathbf{X}^{\mathsf{T}}\mathbf{LX}).
    \label{eqn:smoothness_time_varying}
\end{equation}
Equation (\ref{eqn:smoothness_time_varying}) sums all the Laplacian quadratic form for each $1<t<M$, \ie there is not a temporal relationship between graph signals in different times $t$.

Qiu \etal \cite{qiu2017time} introduced the temporal difference operator $\mathbf{D}_h \in \mathbb{R}^{M \times (M-1)}$ with the purpose of including temporal information in the problem of time-varying graph signal reconstruction such that:
\begin{equation}
\mathbf{D}_h = 
    \begin{bmatrix} 
    -1 &    &        &    \\
     1 & -1 &        &    \\
       &  1 & \ddots &    \\
       &    & \ddots & -1 \\
       &    &        &  1 \\
    \end{bmatrix}
    \in \mathbb{R}^{M \times (M-1)},
    \label{eqn:temporal_difference_operator}
\end{equation}
and the temporal difference signal as:
\begin{equation}
    \mathbf{XD}_h = [\mathbf{x}_2-\mathbf{x}_1,\mathbf{x}_3-\mathbf{x}_2,\dots,\mathbf{x}_M-\mathbf{x}_{M-1}].
    \label{eqn:difference_signal}
\end{equation}
Qiu \etal \cite{qiu2017time} also proposed two time-varying graph signal reconstruction batch methods. The first method is focused in the noiseless case, while the second method approaches the noisy case. In this paper we focus in the noisy case defined as follows:
\begin{equation}
    \min_{\mathbf{\hat{X}}} \frac{1}{2} \Vert \mathbf{J} \circ \mathbf{\mathbf{\hat{X}}} - \mathbf{Y} \Vert_F^2 + \frac{\lambda}{2} \tr\left((\mathbf{\hat{X}D}_h)^{\mathsf{T}}\mathbf{L\hat{X}D}_h\right),
    \label{eqn:qiu_reconstruction_noisy}
\end{equation}
where $\mathbf{J} \in \{0,1\}^{N \times M}$ is the sampling matrix for the whole time-varying graph signal, and $\mathbf{Y} \in \mathbb{R}^{N\times M}$ is the matrix of observed values (the data that we know). $\mathbf{J}$ is defined as follows:
\begin{equation}
    \mathbf{J}(i,j) = \begin{cases}
        1 & \text{if } i \in \mathcal{S}_t, \\
        0 & \text{if } i \notin \mathcal{S}_t,
    \end{cases}
\end{equation}
where $\mathcal{S}_t$ is the set of sampled nodes at time $t$. Basically, Eqn. (\ref{eqn:qiu_reconstruction_noisy}) is trying to reconstruct a time-varying graph signal $\mathbf{\hat{X}}$ with a small error $\Vert \mathbf{J} \circ \mathbf{\mathbf{\hat{X}}} - \mathbf{Y} \Vert_F^2$ while minimizing the smoothness of the temporal difference graph signal.

\subsection{Graph Construction}

Let $\mathbf{M} \in \mathbb{R}^{N\times 2}$ be the matrix of locations of all nodes in $\mathcal{V}$ such that $\mathbf{M} = [\mathbf{m}_1,\dots,\mathbf{m}_N]^{\mathsf{T}}$, where $\mathbf{m}_i \in \mathbb{R}^2$ is the vector with latitude and longitude of node $i$. A k-nearest neighbors algorithm with $\text{k}=10$ is used to connect the nodes in the graph for the experiments in this paper. The weight of the edge $(i,j)$ is such that $\mathbf{W}(i,j) = \exp{({-\frac{d(i,j)^2}{\sigma^2}})}$, where $d(i,j)=\Vert \mathbf{m}_i - \mathbf{m}_j \Vert_2$ is the euclidean distance between nodes $(i,j)$, and $\sigma^2$ is the standard deviation of the Gaussian function computed as follows:
\begin{equation}
    \sigma = \frac{1}{\vert \mathcal{E} \vert + N}\sum_{(i,j) \in \mathcal{E}} d(i,j).
    \label{eqn:standard_deviation}
\end{equation}

\subsection{Sobolev Norm Reconstruction of Time-Varying Graph Signals}
\label{sec:sobolev_norm}

In this paper, we propose a new time-varying graph signals reconstruction algorithm inspired by the minimization of the Sobolev norm in GSP. The definition of this norm was given by Pesenson \etal \cite{pesenson2009variational}, who used the Sobolev norm to introduce the variational problem in graphs.
\begin{definition}
    \label{dfn:sobolev_norm}
    For a fixed $\epsilon \geq 0$, the Sobolev norm is defined as follows:
    \begin{equation}
        \Vert \mathbf{f} \Vert_{\beta,\epsilon} = \Vert (\mathbf{L}+\epsilon \mathbf{I})^{\beta/2} \mathbf{f} \Vert, \beta \in \mathbb{R}.
        \label{eqn:sobolev_norm}
    \end{equation}
\end{definition}
When $\mathbf{L}$ is symmetric, Eqn. (\ref{eqn:sobolev_norm}) can be rewritten as follows:
\begin{equation}
    \mathbf{f}^{\mathsf{T}}(\mathbf{L}+\epsilon\mathbf{I})^{\beta}\mathbf{f}
    \label{eqn:sobolev_norm_rewritten}
\end{equation}

Giraldo and Bouwmans \cite{giraldo2020graphbgs} found that the term $(\mathbf{L}+\epsilon\mathbf{I})$ in the Sobolev norm in Eqn. (\ref{eqn:sobolev_norm}) has a better condition number than $\mathbf{L}$, they used the following theorems:
\begin{theorem}
    \label{trm:perturbation_matrix}
    Let $\mathbf{\Psi} \in \mathbb{R}^{N\times N}$ be a perturbation matrix. Given a combinatorial Laplacian matrix $\mathbf{L}$, the term $\mathbf{L} + \mathbf{\Psi}$ has a lower and upper bound in the condition number such that:
    \begin{equation}
        \frac{\sigma_{\text{max}}(\mathbf{L}+\mathbf{\Psi})}{\sigma_{\text{max}}(\mathbf{\Psi})} \leq \kappa(\mathbf{L}+\mathbf{\Psi}) \leq \frac{\sigma_{\text{max}}(\mathbf{L})+\sigma_{\text{max}}(\mathbf{\Psi})}{\sigma_{min}(\mathbf{L}+\mathbf{\Psi})},
        \label{eqn:condition_number}
    \end{equation}
    where $\kappa(\mathbf{L}+\mathbf{\Psi})$ is the condition number of $\mathbf{L}+\mathbf{\Psi}$.\\
    Proof: see \cite{giraldo2020graphbgs}.
\end{theorem}

\begin{theorem}
    \label{trm:weyl_theorem}
    Let $\mathbf{L}$ and $\mathbf{\Psi}$ be Hermitian matrices with set of eigenvalues $\{\lambda_1,\dots,\lambda_N\}$ and $\{\psi_{1},\dots,\psi_{N}\}$, respectively. The matrix $\mathbf{L} + \mathbf{\Psi}$ has a set of eigenvalues $\{\nu_{1},\dots,\nu_{N}\}$ where the following inequalities hold for $i=1,\dots,N$:
    \begin{equation}
        \lambda_i+\psi_1 \leq \nu_i \leq \lambda_i + \psi_N.\\
    \end{equation}
    Proof: see \cite{horn2012matrix}.
\end{theorem}
From Theorem \ref{trm:perturbation_matrix}, we can notice that the condition number of $(\mathbf{L}+\epsilon\mathbf{I})$ is lower bounded by $\sigma_{\text{max}}(\mathbf{L}+\epsilon\mathbf{I})/\epsilon$, \ie greater values of $\epsilon$ could end up in better condition numbers. In the same way, we have that $\kappa(\mathbf{L}+\epsilon\mathbf{I}) < \infty$ since $\sigma_{min}(\mathbf{L}+\mathbf{\Psi})$ is strictly greater than zero according to Theorem \ref{trm:weyl_theorem}, while $\kappa(\mathbf{L}) = \infty$ because the first eigenvalue of $\mathbf{L}$ is zero, \ie we are improving the condition number of $\mathbf{L}$ when adding the perturbation matrix $\epsilon\mathbf{I}$. In this paper, we use this fact to show that the convergence rate of the Sobolev norm is better than the graph Laplacian quadratic form when solving with gradient descent.

Firstly, we need to extend the definition of the Sobolev norm to time-varying graph signals as follows (given that $\mathbf{L}$ is a symmetric matrix):
\begin{equation}
    \Vert \mathbf{X} \Vert_{\beta,\epsilon} = \sum_{i=1}^M \mathbf{x}_i^{\mathsf{T}}(\mathbf{L}+\epsilon\mathbf{I})^{\beta}\mathbf{x}_i = \tr(\mathbf{X}^{\mathsf{T}}(\mathbf{L}+\epsilon\mathbf{I})^{\beta}\mathbf{X}).
    \label{eqn:sobolev_norm_time}
\end{equation}
The Sobolev reconstruction problem for time-varying graph signals is formulated as follows:
\begin{equation}
    \min_{\mathbf{\hat{X}}} \frac{1}{2} \Vert \mathbf{J} \circ \mathbf{\mathbf{\hat{X}}} - \mathbf{Y} \Vert_F^2 + \frac{\lambda}{2} \tr\left((\mathbf{\hat{X}D}_h)^{\mathsf{T}}(\mathbf{L}+\epsilon\mathbf{I})^{\beta}\mathbf{\hat{X}D}_h\right),
    \label{eqn:sobolev_reconstruction_noisy}
\end{equation}
where we used the temporal difference operator in Eqn. (\ref{eqn:temporal_difference_operator}), and the Sobolev norm of time-varying graph signals in Eqn. (\ref{eqn:sobolev_norm_time}). Equation (\ref{eqn:sobolev_reconstruction_noisy}) is solved with conjugate gradient method in this paper.

\subsection{Rate of Convergence}
\label{sec:rate_convergence}

The rate of convergence of the Sobolev norm reconstruction in Eqn. (\ref{eqn:sobolev_reconstruction_noisy}) is better than the rate of convergence of the method involving the graph Laplacian quadratic form in Eqn. (\ref{eqn:qiu_reconstruction_noisy}). Intuitively, we are improving the condition number of the Sobolev norm with respect to the Laplacian quadratic form as shown in Section \ref{sec:sobolev_norm}. As a consequence, one can expect that the Sobolev reconstruction problem is better conditioned, and then a gradient descent method can arrive faster to the global minimum (when the optimization problem is convex).

Formally, the rate of convergence of a gradient descent method is at best linear. This rate can be accelerated if the condition number of the Hessian of the cost function is reduced \cite{arora2017more}. Qiu \etal \cite{qiu2017time} showed that the problem in Eqn. (\ref{eqn:qiu_reconstruction_noisy}) can be rewritten as follows:
\begin{equation}
    \min_{\mathbf{z}} \frac{1}{2} \Vert \mathbf{Q}[\mathbf{z}-\vectorized{(\mathbf{Y})}] \Vert_2^2+\frac{\lambda}{2}\mathbf{z}^{\mathsf{T}}[(\mathbf{D}_h\mathbf{D}_h^{\mathsf{T}})\otimes \mathbf{L}] \mathbf{z} = f(\mathbf{z}),
    \label{eqn:qiu_reconstruction_noisy_rewritten}
\end{equation}
where $\mathbf{Q}=\diag(\vectorized{(\mathbf{J})})\in \mathbb{R}^{MN\times MN}$, and $\mathbf{z}=\vectorized{(\mathbf{\hat{X}})}$. In the same way, the Hessian matrix of $f(\mathbf{z})$ in Eqn. (\ref{eqn:qiu_reconstruction_noisy_rewritten}) is such that:
\begin{equation}
    \nabla_{\mathbf{z}}^2 f(\mathbf{z}) = \mathbf{Q} + \lambda (\mathbf{D}_h\mathbf{D}_h^{\mathsf{T}})\otimes \mathbf{L}.
    \label{eqn:hessian_quadratic_form}
\end{equation}
Using a similar rationale, the Hessian matrix associated with the Sobolev reconstruction problem is such that:
\begin{equation}
    \nabla_{\mathbf{z}}^2 f(\mathbf{z}) = \mathbf{Q} + \lambda (\mathbf{D}_h\mathbf{D}_h^{\mathsf{T}})\otimes (\mathbf{L}+\epsilon\mathbf{I})^{\beta}.
    \label{eqn:hessian_sobolev}
\end{equation}
If we analyze the condition number of the second term in Eqn. (\ref{eqn:hessian_quadratic_form}) we get:
\begin{equation}
    \kappa(\lambda (\mathbf{D}_h\mathbf{D}_h^{\mathsf{T}})\otimes \mathbf{L}) = \kappa(\mathbf{D}_h\mathbf{D}_h^{\mathsf{T}}) \kappa(\mathbf{L}),
\end{equation}
where we applied the following property \cite{xiang2005perturbation}:
\begin{equation}
    \kappa(\mathbf{A}\otimes \mathbf{B})\vcentcolon= \Vert \mathbf{A}^{-1} \Vert \Vert \mathbf{A} \Vert \Vert \mathbf{B}^{-1} \Vert \Vert \mathbf{B} \Vert = \kappa(\mathbf{A})\kappa(\mathbf{B}).
\end{equation}
Similarly, the condition number of the second term in Eqn. (\ref{eqn:hessian_sobolev}) is:
\begin{equation}
    \kappa(\lambda (\mathbf{D}_h\mathbf{D}_h^{\mathsf{T}})\otimes (\mathbf{L}+\epsilon\mathbf{I})^{\beta}) = \kappa(\mathbf{D}_h\mathbf{D}_h^{\mathsf{T}}) \kappa((\mathbf{L}+\epsilon\mathbf{I})^{\beta}).
\end{equation}
Then for $\beta=1$ and $\epsilon>0$, we know from Section \ref{sec:sobolev_norm} that $\kappa(\mathbf{L}+\epsilon\mathbf{I}) < \kappa(\mathbf{L})$ and then:
\begin{equation}
    \kappa(\mathbf{D}_h\mathbf{D}_h^{\mathsf{T}}) \kappa(\mathbf{L}+\epsilon\mathbf{I}) < \kappa(\mathbf{D}_h\mathbf{D}_h^{\mathsf{T}}) \kappa(\mathbf{L}).
\end{equation}
As a consequence, we have that:
\begin{equation}
    \kappa(\lambda (\mathbf{D}_h\mathbf{D}_h^{\mathsf{T}})\otimes (\mathbf{L}+\epsilon\mathbf{I})) < \kappa(\lambda (\mathbf{D}_h\mathbf{D}_h^{\mathsf{T}})\otimes \mathbf{L}).
\end{equation}
In other words, we are improving the convergence rate of the Sobolev reconstruction with respect to the problem with the Laplacian quadratic form by reducing the condition number of the Hessian associated to these optimization problems.

\section{Experimental Framework}
\label{sec:experimental_framework}

This section introduces the databases used in this paper, and the experiments to validate our proposed method.

\subsection{Databases}

\begin{figure}
    \centering
    \includegraphics[width=0.48\textwidth]{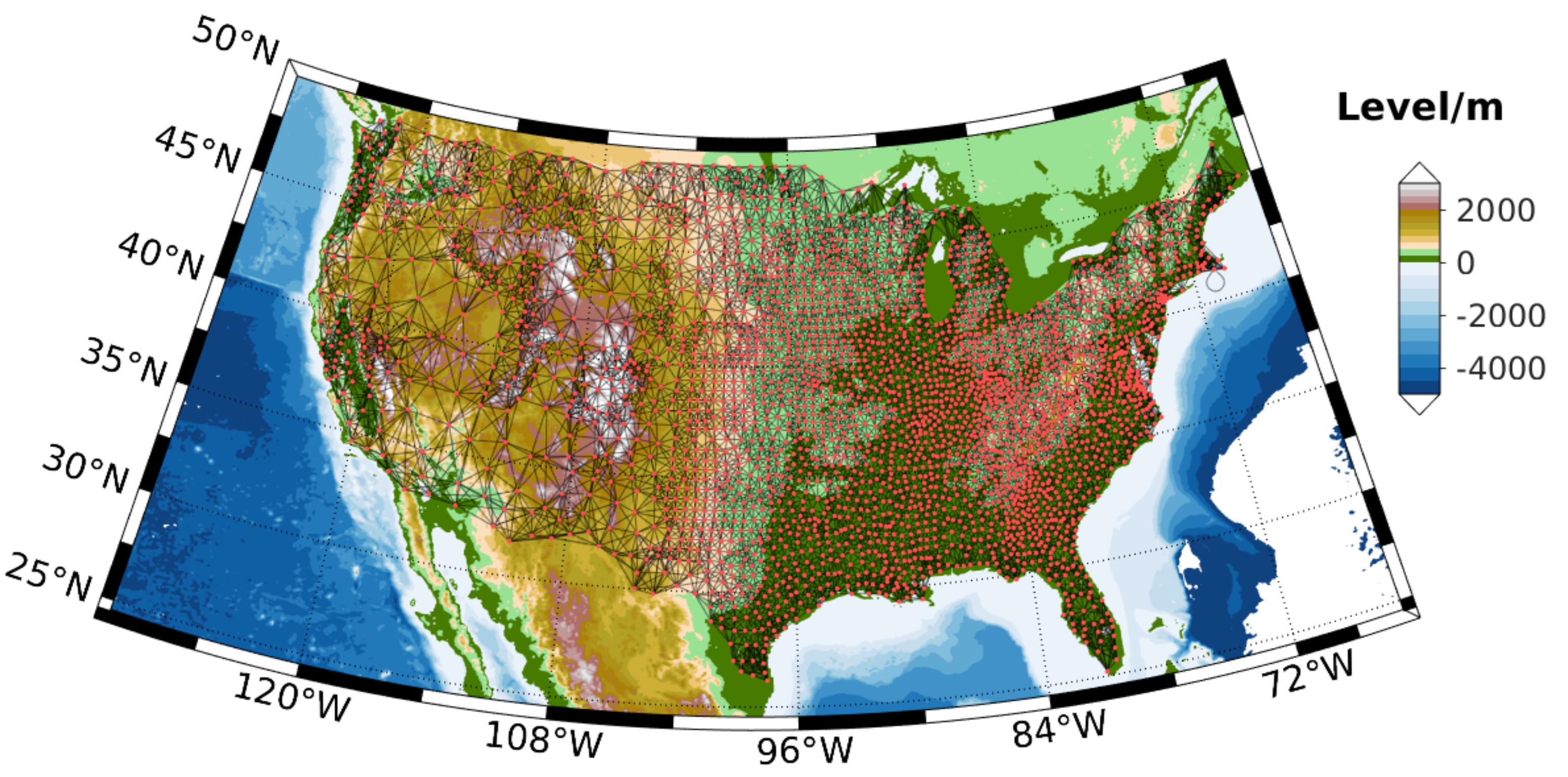}
    \caption{Graph with the cities in the United States in the Johns Hopkins University database \cite{dong2020interactive}. This graph was constructed with k-nearest neighbors with $\text{k}=10$.}
    \label{fig:usa_covid_cases}
\end{figure}

We use the global and the USA COVID-19 databases provided by the Johns Hopkins University \cite{dong2020interactive}. The databases were used between the dates January 22, 2020 and April 6, 2020. The global dataset contains the cumulative number of COVID-19 cases for each day and each locality between January 22 and April 6, as well as the geographical localization of 259 places including some regions of the world (Figure \ref{fig:global_covid_cases} shows the localities in the database). In the same way, the USA database contains information of 3149 localities about COVID-19 in the United States. Figure \ref{fig:usa_covid_cases} shows the graph $G$ used in this paper for the USA COVID-19 database. The sampling period of both databases is 1 day. We pre-processed the COVID-19 databases to get the number of new cases each day instead of the cumulative number of cases.

We also use a database of sea surface temperature, and a dataset of daily mean concentration of Particulate Matter (PM) 2.5 in California for reference purposes. These databases were used in the paper of Qiu \etal \cite{qiu2017time}.

\subsection{Experiments}

We compare our Sobolev method against Qiu's algorithm \cite{qiu2017time}, and Natural Neighbor Interpolation (NNI) \cite{sibson1981brief}. For each database, we make one experiment to check the performance in all methods, and one additional experiment to check the convergence rate of our Sobolev algorithm and Qiu's method. In the first experiment, for Qiu's method we first search the best $\lambda$ parameter in Eqn. (\ref{eqn:qiu_reconstruction_noisy}) by performing $5$ reconstruction for each $\lambda$ in the set $\mathcal{M}=\{1\times 10^{-3}, 1\times 10^{-2}, 2\times 10^{-2}, 5\times 10^{-2}, 0.1, 0.2, 0.5, 1, 2, 5, 10,$ $20, 50, 1\times 10^{2}, 2\times 10^{2}, 5\times 10^{2}\}$, and with sampling densities: $\{0.1, 0.2,\dots, 0.9\}$ for the databases of sea surface temperature and PM 2.5, and $\{0.5, 0.6, \dots, 0.9, 0.995\}$ for the COVID-19 datasets. We use a random sampling strategy for $\mathbf{J}$, ensuring that each time graph signal $\mathbf{x}_i$ has the same amount of sampled nodes for all $1<i<M$. Secondly, we execute the algorithm with the best $\lambda$ parameter $100$ times (with different $\mathbf{J}$ matrices) for each sampling density. For our Sobolev method in Eqn. (\ref{eqn:sobolev_reconstruction_noisy}) we use the same strategy, but in this case the best parameters are searched in the cartesian product $\mathcal{M}\times \mathcal{M}$ for $\lambda \in \mathcal{M}$ and $\epsilon \in \mathcal{M}$. Finally, for NNI we perform the experiment $100$ times for each sampling density.

The second experiment records the number of iterations required to converge for the reconstruction method in Eqn. (\ref{eqn:qiu_reconstruction_noisy}) and in Eqn. (\ref{eqn:sobolev_reconstruction_noisy}). This experiment is repeated $100$ times for each sampling density using the best parameters $\lambda$ and $\epsilon$ found in the first experiment for the Sobolev method. For a fair comparison, both reconstruction methods use the same $\mathbf{J}$ and $\lambda$ in each repetition.

\section{Results and Discussion}
\label{sec:results_and_discussion}

\begin{figure*}
    \centering
    \includegraphics[width=\textwidth]{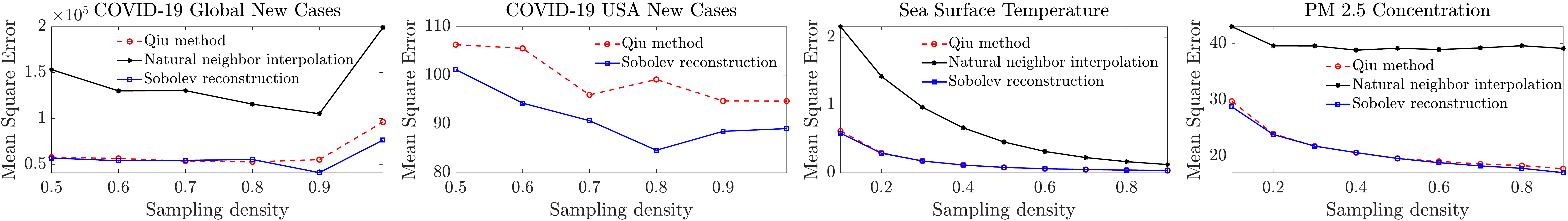}
    \caption{Average mean square error using the reconstruction method proposed by Qiu \etal in \cite{qiu2017time}, natural neighbor interpolation \cite{sibson1981brief}, and our method with Sobolev norm, versus the sampling density considering the reconstruction in $4$ databases with $100$ different sampling matrices $\mathbf{J}$.}
    \label{fig:results_performance}
\end{figure*}

\begin{figure*}
    \centering
    \includegraphics[width=\textwidth]{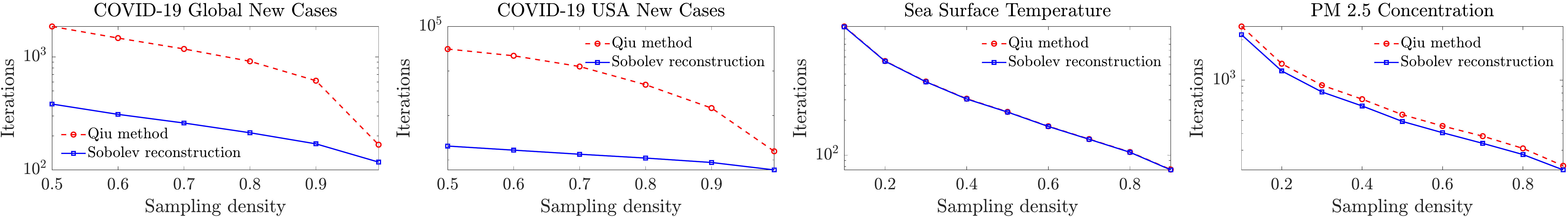}
    \caption{Average number of iterations in logarithmic scale in the y-axis, using the reconstruction method proposed by Qiu \etal in \cite{qiu2017time} and our method with Sobolev norm, versus the sampling density considering the reconstruction in $4$ databases with $100$ different sampling matrices $\mathbf{J}$.}
    \label{fig:results_iterations}
\end{figure*}

Figure \ref{fig:results_performance} shows the average mean square error in the four databases using Qiu's method \cite{qiu2017time}, NNI \cite{sibson1981brief}, and our Sobolev algorithm. Figure \ref{fig:results_performance} shows that our method is better than Qiu's algorithm and NNI in both global and USA COVID-19 datasets. The results of NNI are not displayed in the USA COVID-19 database because its performance is very far from Qiu and Sobolev methods. Our method also outperforms NNI in the sea surface temperature and PM 2.5 concentration, while having relatively the same performance with respect to Qiu's algorithm. Moreover, Figure \ref{fig:results_iterations} shows the average number of iterations with both Qiu's method and our proposed Sobolev algorithm, in this case the Sobolev reconstruction clearly performs better than Qiu's method for the COVID-19 databases, and slightly better in the PM 2.5 concentration dataset. The behavior in the number of iterations of both methods in Figure \ref{fig:results_iterations} is expected according to the proofs in Section \ref{sec:rate_convergence}. Then, we can argue that our Sobolev method is better for time-varying graph signals reconstruction than Qiu's method \cite{qiu2017time} for the estimation of new COVID-19 cases.

It is also important to remark that the estimation of new COVID-19 in the USA dataset shows promising results. This algorithm could be useful in two scenarios. Firstly, we can use the Sobolev norm minimization to estimate the number of possible COVID-19 cases in certain regions without confirmed cases, \ie we can add a node in $G$ with the desired location and perform the Sobolev reconstruction. Secondly, the minimization of the Sobolev norm could be incorporated for leveraging spatio-temporal information in well-established models of infectious diseases \cite{daley2001epidemic,hethcote2000mathematics}, which we leave for future work.

\section{Conclusions}
\label{sec:conclusions}

This paper introduces an estimation method of new COVID-19 cases. This estimation is performed using spatio-temporal data embedded in a time-varying graph signal. Our method is based in the minimization of the Sobolev norm, and outperforms state-of-the-art algorithms in time-varying graph signals reconstruction. Most importantly, we show the benefits of the convergence rate of our method by relying on the condition number of the Hessian associated with the problem.

The present work opens up several questions for future research. For example, what is the role of GSP in the field of the mathematical modeling of infectious disease? Or, what is the importance of the Sobolev norm for the acceleration of the convergence rate in the underlying optimization problems of mainstream applications in GSP such as graph convolutional neural networks \cite{kipf2016semi,bronstein2017geometric}, active learning \cite{gadde2014active,anis2018sampling}, among others,

\bibliographystyle{IEEEbib}
\bibliography{bibfile}

\end{document}